\documentclass[aps,twocolumn,nofootinbib,superscriptaddress,preprintnumbers,floatfix]{revtex4}
\usepackage[bookmarks=false]{hyperref}
\usepackage{graphicx}
\usepackage{epsfig,amsmath,euscript}

\newcommand{\be}{\begin{equation}}
\newcommand{\ee}{\end{equation}}

\def\d{{\rm d}}
\def\OMIT#1{{}}

\newcommand{\mcdot}{\!\cdot\!}

\newcommand{\bra}[1]{\left\langle #1\right\rvert}
\newcommand{\ket}[1]{\left\lvert #1\right\rangle}

\newcommand{\e}{\mathrm{e}}

\newcommand{\eq}[1]{Eq.~\eqref{#1}}
\newcommand{\vc}[1]{\boldsymbol{#1}}

\begin{document}

\title{Electroweak radiative corrections and unitarity of Standard Model}
\author{Grigory Ovanesyan}
\affiliation{Los Alamos National Laboratory, Theoretical Division, MS B238, Los Alamos, NM 87545, USA}
\begin{abstract}
Electroweak corrections to longitudinal gauge and Higgs boson scattering amplitudes are calculated. Due to Sudakov double logarithms, the effect is a suppression of amplitude that grows rapidly with increasing center of mass energy leading to significant reduction of cross sections compared to tree level results. For example, the suppression factor for the cross section of $W^+_{\text{L}}W^-_{\text{L}}$ scattering due to these corrections varies from $0.3$ at the center of mass energy of $100\text{TeV}$ to a factor of $10^{-12}$ at the center of mass energy of $10^{13}\text{TeV}$. The modification of SM unitarity bound due to these corrections is obtained.
\end{abstract} 
\maketitle

\section{Introduction}\label{introsec}
It is well known that in Standard Model (SM) the Higgs quartic coupling has a Landau pole for a sufficiently heavy Higgs mass. The position of the pole is a function of the Higgs boson mass. For light enough mass of the Higgs boson, either Landau pole is higher than the Plank scale or it disappears, and the unitarity of SM is preserved. However very quickly as $m_H$ approaches $~700\text{GeV}$ the Landau pole becomes close to the $\text{TeV}$ scale and as a consequence the perturbative unitarity is broken in longitudinal gauge boson scattering amplitudes. This is known as unitarity bound on Higgs boson mass \cite{Lee:1977eg,Marciano:1989ns, Kolda:2000wi}, which relates the Higgs mass to the energy scale at which the unitarity is broken in SM. This scale, as a function of $m_H$ represents a scale at which either some new particles should come in and save unitarity, or otherwise, the SM becomes strongly coupled. Note that in this paper we refer to unitarity of SM in the presence of the Higgs boson, not to be confused with analysis of longitudinal gauge boson scattering in the absence of Higgs boson, in which case unitarity is violated at much lower scales.

In the standard analysis of unitarity bound on Higgs mass for SM the equivalence theorem\cite {Lee:1977eg,Chanowitz:1985hj} is used to relate the longitudinal gauge boson scattering amplitude to the corresponding scalar scattering, including unphysical Goldstone boson modes in the $R_{\xi}$ gauge. At tree level and high energies these amplitudes are purely $s-$waves, proportional to Higgs self-coupling $\lambda=G_F m_H^2/\sqrt{2}$.

Naive analysis of loop contributions of SU(2) and U(1) gauge bosons to say $W_LW_L\rightarrow W_L W_L$ amplitude seems to give very small result, because couplings $\alpha_2=\alpha_{\text{em}}/\sin^2\theta_W\approx 0.03, \alpha_1=\alpha_{\text{em}}/\cos^2\theta_W\approx 0.01$ are small. However such amplitudes contain Sudakov double logarithms\cite{Denner:2001gw}-\cite{Denner:2003wi}, and in a series of recent papers \cite{Chiu:2007yn}$-$\cite{Fuhrer:2010eu} it was shown how in the effective theory approach one can resum  these large logarithms. The results of  \cite{Chiu:2007yn}$-$\cite{Fuhrer:2010eu} are that at the $\text{TeV}$ scale such Sudakov resummation leads to decrease in the cross section for processes like $q\bar{q}\rightarrow W_T W_T$ up to $40\%$, a suppression factor that rapidly increases with the center of mass energy. In a recent paper \cite{Manohar:2012rs} it was shown that electroweak Sudakov corrections play a significant role in top quark forward backward asymmetry.

It is the purpose of this paper to go beyond the traditional tree level unitarity analysis of SM and study how higher order corrections, which are part of SM and can be calculated using existing techniques, modify the unitarity bound of SM. The paper is organized as follows: in section \ref{seq:theo} we briefly review the approach developed in \cite{Chiu:2007yn}$-$\cite{Fuhrer:2010eu} to resum electroweak logarithms for high-energy processes. In section \ref{seq:WW} we derive the effect of electroweak large logarithm resummation on the high energy behavior of longitudinal gauge bosons and Higgs scattering amplitudes. We study the applications of derived corrections to the SM unitarity bound in section \ref{seq:Unitarity}. We conclude in section \ref{seq:concl} .

\section{Theoretical Framework}\label{seq:theo}
Soft collinear effective theory (SCET) \cite{Bauer:2000ew}$-$\cite{Bauer:2001ct} is an effective theory for highly energetic quarks and gluons. Recently it has been extended in \cite{Chiu:2007yn}$-$\cite{Fuhrer:2010eu} to Standard Model to compute renormalization group improved amplitudes for SM including resummation of Sudakov large double-logarithms from loops with SU(2) and U(1) gauge bosons. The computation can be divided in three steps.

First we integrate out the hard modes at the scale of center of mass energy $\mu_h=\sqrt{s}$ for our scattering process. For $2\rightarrow2$ process we match on four-particle contact operators in EFT, which is referred to as $\text{SCET}_{\text{EW}}$ with massless SU(2) and U(1) gauge bosons. This leaves us with some basis of operators and Wilson coefficients, which only contain $\ln{\mu_h}/{\sqrt{s}}$ and can be minimized by scale choice $\mu_h=\sqrt{s}$.

Next the effective operators in $\text{SCET}_{\text{EW}}$ have to be evolved down to the weak scale $M_{\text{EW}}\approx M_z$. In order to relate the Wilson coefficients at low scale to the ones obtained by matching  on previous step at high scale one needs to exponentiate the anomalous dimension matrix, which consists from collinear and soft parts. This anomalous dimension has been studied and calculated to one loop order for any process in SM in \cite{Chiu:2007yn}$-$\cite{Fuhrer:2010eu} . 

Finally at the low scale $\mu_l=M$ one has to match onto broken theory where gauge bosons have been removed. As a result one gets the effective amplitude for scattering of gauge bosons and quarks and leptons with resummed large logarithms. The corresponding amplitude in $\text{SCET}_{\text{EW}}$ gets the following factorized form\cite{Chiu:2009mg}:
\begin{eqnarray}
&&\mathcal{M}=\exp\left[\vc{D}(\alpha(\mu_l))\right] \,P\exp\left(\int_{\mu_h}^{\mu_l}\frac{d\mu}{\mu}\vc{\gamma}(\alpha(\mu))\right)\nonumber\\
&&\qquad\,\,\,\times\vc{c}\left(\alpha(\mu_h),\left\{\ln\frac{p_i\mcdot p_j}{\mu_h^2}\right\}\right),\label{mresummaster}
\end{eqnarray}
where $\vc{c}$ is the column vector of Wilson coefficients obtained from matching at high scale, $\gamma$ is the anomalous dimension of  effective operator, which is a matrix in the color space, and finally $\vc{D}$ appears as a result of the low scale matching. The anomalous dimension for a general hard scattering process has the form $\vc{\gamma}=\gamma_C\vc{1}+\vc{\gamma}_S$, where collinear anomalous dimension is a sum of collinear anomalous dimensions associated with each leg and soft anomalous dimension has a non-trivial color structure but is universal in the sense that it only depends on light-cone directions and color flow.
\section{Longitudinal bosons scattering at high energies}\label{seq:WW}
We derive in this section electroweak radiative correction contribution to the amplitude of longitudinal gauge and/or Higgs boson scattering amplitude. The corresponding formalism has been developed in \cite{Chiu:2007yn}$-$\cite{Fuhrer:2010eu} and was briefly reviewed in section \ref{seq:theo}\, . Following these references we use the equivalence theorem to relate gauge boson amplitude to the corresponding unphysical scalar amplitude in $R_{\xi}$ gauge. This method was applied in Refs. \cite{Chiu:2007yn}$-$\cite{Fuhrer:2010eu} to derive longitudinal gauge boson production at LHC to next-to-next-to-leading logarithmic order(NNLL), except for scalar contribution at two loops that is missing for some parts. We work at next-to-leading logarithmic order (NLL).
\subsection{High-scale matching}
At the scale $\mu_h\sim\sqrt{s}$ we integrate out hard modes and leave only collinear modes, corresponding to four external label momenta. For our purpose the Higgs quartic coupling gives the biggest contribution to the matching. We neglect the gauge boson mediated interactions between the scalars, since they are suppressed by the ratio of weak coupling to Higgs self-coupling $\lambda$.  The tree level matching is given by Higgs quartic term in the SM Lagrangian, and is trivial: 
\begin{eqnarray}
&&O_1=\phi^{\dagger}_4 t^a\phi_3\,\phi^{\dagger}_2t^a\phi_1,\qquad \,\,C_1=0,\label{op1def}\\
&&O_2=\phi^{\dagger}_4 \phi_3\,\phi^{\dagger}_2\phi_1,\qquad\qquad C_2=-\lambda(\mu_h).\label{op2def}
\end{eqnarray}
The Higgs quartic coupling has a familiar Landau pole at higher energies, and is given at one loop order:
\begin{eqnarray}
\lambda(\mu)=\frac{\lambda(m_H)}{1-\frac{3\lambda(m_H)}{2\pi^2}\,\ln(\mu/m_H)},\label{quarticrunning}
\end{eqnarray}
while $\lambda(m_H)=G_F m_H^2/\sqrt{2}$. Equation \eq{quarticrunning} is valid approximation for $m_H>200\text{GeV}$. For lower masses one has to include contribution of top quark Yukawa coupling and the Landau pole disappears. We come back to this case at the end of section \ref{seq:Unitarity}.
\subsection{Running from $\mu_h$ to $\mu_l$}
From high scale the Wilson coefficients must be evolved to the low scale $\mu_l$ which is of the order of the electroweak scale $\mu_l\sim M_W, M_Z$. This is achieved by calculating the anomalous dimension of the four particle operator in the effective theory $\text{SCET}_{\text{EW}}$\cite{Chiu:2009mg}. The anomalous dimension for such process can be written as sum of process dependent collinear anomalous dimension, which is simply the sum over the corresponding terms for each leg, and the universal soft anomalous dimension \cite{Chiu:2008vv, Chiu:2009mg}, which only depends on kinematics and color structure, but is same for say scalars or fermions. Thus the collinear part of anomalous dimension we know from \cite{Chiu:2007yn}$-$\cite{Fuhrer:2010eu} and the soft part of the anomalous dimension is same as in $q\bar{q}\rightarrow q\bar{q}$ for $SU(2)$ and $U(1)$. For $SU(3)$ the soft anomalous dimension to one loop is known \cite{Kidonakis:1998nf}, It was derived for $SU(3)$, $SU(2)$ and $U(1)$ in Refs. \cite{Chiu:2008vv,Chiu:2009mg,Chiu:2009ft}. Since our effective operators are (SU(3))color singlets, the anomalous dimension gets contribution only from SU(2) and U(1) gauge bosons. The result for the total anomalous dimension is:
\begin{eqnarray}
&&\gamma=4\gamma_{\phi}\vc{1}+\gamma_S,\label{gammatot}\\
&&\gamma_{\phi}=\left(\frac{3}{4}\frac{\alpha_2}{4\pi}+\frac{1}{4}\frac{\alpha_1}{4\pi}\right)
\left(2\ln{\frac{s}{\mu^2}}-4\right),\\
&&\gamma_S=\frac{\alpha_2}{\pi}\left(-\frac{3}{2}i\pi\vc{1}+\left[ \begin{array}{cc}
T+U & 2(T-U)  \\
\frac{3}{8}(T-U) & 0 \end{array} \right]\right)\nonumber\\
&&+\frac{\alpha_1}{\pi}2Y_{\phi}^2
\left((T-U)-i\pi\right),\label{gammasoft}
\end{eqnarray}
where the Mandelstam invariants are defined according to $s=(p_1+p_2)^2, t=(p_1-p_4)^2, u=(p_1-p_3)^2$ and 
\begin{eqnarray}
&&T=\ln\frac{-t}{s}+i\pi=\ln\frac{1-\cos\theta}{2}+i\pi,\\
&&U=\ln\frac{-u}{s}+i\pi=\ln\frac{1+\cos\theta}{2}+i\pi,
\end{eqnarray}
where $\cos\theta$ is the scattering angle in the center of mass frame, defined as angle between momenta of $p_1$ and $p_4$ in the CM frame. We will need the formula for the anomalous dimension in the universal form in terms of Mandelstam variables $s,t,u$, without assuming that $s>0, t,u<0$ as in \eq{gammatot} above. The corresponding analytic continuation looks as follows:
\begin{eqnarray}
&&\gamma(\mu, s, t, u) = \frac{\alpha_1}{\pi}\left(\frac{1}{2}\ln\frac{-s}{\mu^2}-1+\frac{1}{2}\text{L}_{t/u}\right)\nonumber\\
&&+\frac{\alpha_2}{\pi}\left(\frac{3}{2}\left(\ln\frac{-s}{\mu^2}-2\right)+\left[ \begin{array}{cc}
\text{L}_{tu/s^2}& 2\text{L}_{t/u}  \\
\frac{3}{8}\text{L}_{t/u} & 0 \end{array} \right]\right),\label{gtotanalytic}
\end{eqnarray}
where $\text{L}_{t/u}=\ln(-t-i0^+)-\ln(-u-i0^+)$, $\text{L}_{tu/s^2}=\ln(-t-i0^+)+\ln(-u-i0^+)-2\ln(-s-i0^+)$ \cite{Chiu:2008vv}.

It is useful to have an analytical formula for (matrix) exponential of the integral from $\mu_h$ to $\mu_l$ of anomalous dimension in \eq{gtotanalytic} that appears in the resummed amplitude in \eq{mresummaster}:
  \begin{eqnarray}
\Gamma\equiv\left(
\begin{array}{cc}
\Gamma_{11}& \Gamma_{12}  \\
\Gamma_{21}&\Gamma_{22}  \end{array}\right)=P\exp\left(\int_{\mu_h}^{\mu_l}\frac{d\mu}{\mu}\vc{\gamma}(\alpha(\mu))\right).
\end{eqnarray}
Evaluation of $\Gamma$ requires standard tricks to switch integration from $d\mu$ to $\d\alpha$ using the beta-function as well as matrix exponentiation. The former has been analytically performed in the Appendix A of Ref. \cite{Chiu:2009ft} while the latter is a simple exercise for 2 by 2 matrices. Rewriting the anomalous dimension in terms of cusp and non-cusp part:
\begin{eqnarray}\label{anomaloussplit}
\gamma=\sum_{k=1}^{2}\,\left(a_k\, A_{1k}+a_{k}^2\,A_{2k}\right)\ln\frac{\mu}{\mu_h}+a_k\,B_{1k},
\end{eqnarray} 
where $a_k=\alpha_k/4\pi$ and inex $k=1$ corresponds to $U(1)$ and $k=2$ to $SU(2)$ parts of the SM gauge group and values for cusp and non-cusp parts are summarized in the table below. Note that the $\mu_h$ dependence in the non-cusp part of the anomalous dimension is cancelled exactly by same dependence in the log term in \eq{anomaloussplit} and is introduced in order to use master formula from Appendix A of Ref. \cite{Chiu:2009ft} for the integral of the anomalous dimension. Finally we get for the exponential factor in \eq{mresummaster} a fully analytic expression:
\begin{widetext}
\begin{eqnarray}
&&\Gamma_{\text{NLL}}(\mu_h, \mu_l, s, t, u)=\e^{\Omega-w\text{L}_{ut/s^2}/2}\left[\cosh w\delta\,\vc{1}+\frac{\sinh w\delta}{\delta}\left(
\begin{array}{cc}
-\frac{\text{L}_{ut/s^2}}{2}&-2\text{L}_{t/u}  \\
-\frac{3}{2}\text{L}_{t/u}&\frac{\text{L}_{ut/s^2}}{2}  \end{array}\right)\right], \,\,\,\text{where}\\
&&\Omega=\sum_{k=1}^2\frac{{\pi A_{1k}}\Big[z_k\ln z_k+1-z_k\Big]}{b_{0k}^2\,\alpha_k(\mu_l)}+\frac{A_{1k}b_{1k}}{4b_{0k}^3}\left[\ln z_k-z_k-\frac{1}{2}\ln^2{z_k}+1\right]+\frac{A_{2k}}{4b_{0k}^2}\left[z_k-\ln z_k-1\right]-\frac{\tilde{B}_{1k}}{2b_{0k}}\ln z_k,\label{Omegadinasour}\\
&& w=\frac{2}{b_{02}}\ln z_2, \qquad\qquad\delta=\sqrt{\frac{1}{4}\text{L}_{tu/s^2}^2+\frac{3}{4}\text{L}_{t/u}^2}.
\end{eqnarray}
\end{widetext}
In the equation above we defined $z_k=\alpha_k(\mu_l)/\alpha_k(\mu_h)$, $b_{0k}, b_{1k}$ are the two lowest order beta-function coefficients for $U(1)$ and $SU(2)$ and $A_{1k}, A_{2k}, B_{1k}$ are the one and two-loop cusp anomalous dimension coefficients and one-loop non-cusp anomalous dimension coefficient. They are all summarized in the table below. Note that $\tilde{B}_k$ is defined in such a way that it is same as $B_{1k}$ for $k=1$ and is equal to part of $B_{12}$ which is proportional to unit matrix, i.e. omitting term $4U_S$, where $U_S$ is the matrix in the second term of the second line in \eq{gtotanalytic}, which comes from soft anomalous dimension.

The leading-logarithm (LL) expression is significantly simpler and is given by first term in the expression for $\Omega$ in \eq{Omegadinasour}:
\begin{eqnarray}
\Gamma_{\text{LL}}(\mu_h, \mu_l)=\exp\left({\sum_{k=1}^2\frac{{\pi A_{1k}}\Big[z_k\ln z_k+1-z_k\Big]}{b_{0k}^2\,\alpha_k(\mu_l)}}\right),\label{eq:GammaLL}
\end{eqnarray}
and unlike the NLL expression, the LL one is proportional to unit matrix, has no angular dependance at fixed scales $\mu_h, \mu_l$ and does not depend on any momenta ($p_1, p_2,p_3,p_4$) before setting the scale $\mu_h\sim s$.

\begin{tabular}{ |l | c | r | }
  \hline                       
   & k=1 & k=2 \\
  \hline
  $\alpha_k$ & ${\alpha_{\text{em}}}/{\cos^2 \theta_{{W}}}$ & ${\alpha_{\text{em}}}/{\sin^2 \theta_{{W}}}$\\
    $b_{0k}$ & $-{41}/{6}$ & ${19}/{6}$ \\
   $b_{1k}$ & $-{199}/{30}$ & $-35/6$ \\
  $A_{1k}$ & $-4$ & $-12$ \\
  $A_{2k}$ & $(-4)\mcdot(-104)/9$ & $-12\left(\frac{70}{9}-\frac{2\pi^2}{3}\right)$ \\
  $B_{1k}$ & $2\left(\text{L}_{t/u}-\ln\frac{\mu_h^2}{-s}\right)-4$ & $-6\left(\ln\frac{\mu_h^2}{-s}+2\right)+4 U_S$\\
$\tilde{B}_{1k}$ & $2\left(\text{L}_{t/u}-\ln\frac{\mu_h^2}{-s}\right)-4$ & $-6\left(\ln\frac{\mu_h^2}{-s}+2\right)$   \label{tab:table1}\\
  \hline
\end{tabular}

\subsection{Low-scale matching}

At the low scale $\mu_l\sim M_W, M_Z$ we integrate out $W$ and $Z$ bosons, and match onto $\text{SCET}_{\gamma}$ with only photons (and of course gluons, but for our purpose they are irrelevant). At tree level we simply rewrite each doublet $\phi_i$ in our operator basis $O_1, O_2$ in terms of broken fields:
\begin{eqnarray}
\phi=\left(\begin{array}{cc}
iw^{+}  \\
\frac{H-i z}{\sqrt{2}}  \end{array}\right),\label{doubletform}
\end{eqnarray}
where $w^{\pm}={(\varphi^1\mp i\varphi^2)}/\sqrt{2}$ and $z=\varphi^3$ and we omitted the Higgs vev since we are interested in four-particle interactions only. Writing the operators $O_1$ and $O_2$ in terms of four-particle operators involving fields $w^{\pm}, H, z$ is straightforward but contains a great number of terms. It is more convenient to work in the basis of fields $w^{\pm}, \Psi^{\pm}$, where  $\Psi^{\pm}=(H\mp iz)/\sqrt{2}$ since the number of possible terms with such a choice in the effective Lagrangian is minimal. At the low scale $\mu_l$ the operators in \eq{op1def}-\eq{op2def} match onto the following ones:
\begin{eqnarray}
\hat{O}_1&=&w^-_4 w^+_3 w^-_2 w^+_1, \qquad\hat{O}_2=\Psi^-_4 \Psi^+_3 w^-_2 w^+_1,\nonumber\\
\hat{O}_3&=&w^-_4 w^+_3 \Psi^-_2 \Psi^+_1, \qquad\hat{O}_4=\Psi^-_4 \Psi^+_3 \Psi^-_2 \Psi^+_1,\nonumber\\
\hat{O}_5&=&w^-_4 \Psi^+_3 \Psi^-_2 w^+_1, \qquad\hat{O}_6=\Psi^-_4 w^+_3 w^-_2 \Psi^+_1\label{basis}
\end{eqnarray}
with Wilson coefficients $\hat{C}_i$, where
\begin{eqnarray}
 \hat{C}_i= R_{ij} C_j.
\end{eqnarray}
Coefficients $C_j$ at the high scale are given in  \eq{op1def}-\eq{op2def}, while at the low scale they are equal to:
\begin{eqnarray}
C_1(\mu_l)&=&-\lambda(\mu_h)\Gamma_{12},\\
C_2(\mu_l)&=&-\lambda(\mu_h)\Gamma_{22}.
\end{eqnarray}
The matching matrix $R$ at tree level is found simply by substituting \eq{doubletform} into the definition of operators $O_1, O_2$. The result is:
\begin{eqnarray}
R^{(0)}=\left[\begin{array}{cccccc}
\frac{1}{4} & -\frac{1}{4} & -\frac{1}{4} & \frac{1}{4} & \frac{1}{2} & \frac{1}{2} \\
1 & 1& 1& 1 & 0 & 0
\end{array}\right]^T.
\end{eqnarray}
 Finally for consistency if we stay at NLL order we need to include the matching at low scale at one loop, because this matching contains large logarithms $\ln \sqrt{s}/M_W, \ln \sqrt{s}/M_Z$ and due to finite difference $M_W-M_Z$  in SM, these large logarithm has to be included at NLL order in SM \cite{Chiu:2009mg, Chiu:2009ft}. This calculation in SM is tricky and has been performed consistently in Ref. \cite{Chiu:2009ft}. The result is that the scalar doublet has to be matched on physical states below $\mu_l$, which are $W_L, Z_L, H$ and each component of the doublet gets different matching correction which can be found in \cite{Chiu:2009ft} for general case $\mu_l, M_W, M_Z$. For NLL order and setting $\mu_l=M_Z$, the entire doublet gets the same matching correction. The resummed matrix element gets multiplied as a result of low scale matching by a factor $\exp(D)$, where $D$ has simple expression for this case:
\begin{eqnarray}
D(\mu_l=M_Z)=-\frac{\alpha_W(M_Z)}{\pi}\ln\frac{M_Z}{M_W}\ln\frac{s}{M_Z^2}.
\end{eqnarray}
By setting $\mu_l=M_Z$ we get a simple expression, however the price we pay is that we will not be able to evaluate the low-scale variation consistently. We will only include high-scale variation $\mu_h$ in our plots below. Including the low scale matching, the tree level matrix $R^{(0)}$ gets modified:
\begin{eqnarray}
R=R^{(0)}\mcdot \,\e^{D(\mu_l=M_Z)}.
\end{eqnarray}
\subsection{Lagrangian of EFT}
Now that we have calculated the matrix element including resummation of electroweak Sudakov logarithms, it is straightforward to construct the Lagrangian of effective theory:
\begin{eqnarray}
\mathcal{L}_{\text{EFT}}=\sum_{p_1,p_2,p_3,p_4}  \hat{C}_{i}(p_1,p_2,p_3,p_4)\,\hat{O}_i(p_1,p_2,p_3,p_4)
\end{eqnarray}
Another simplification we will use is to rewrite the color-triplet operator as a combination of two color singlet operators with different contractions:
\begin{eqnarray}
O_1=\frac{1}{2}\phi_4^{\dagger}\phi_1 \, \phi_2^{\dagger}\phi_3-\frac{1}{2N}\phi_4^{\dagger}\phi_3 \, \phi_2^{\dagger}\phi_1,\label{identity}
\end{eqnarray}
with $N=2$ for $SU(2)$. Thus, since momenta $p_1, p_2, p_3, p_4$ have to be summed over, and also the Lagrangian before low-scale matching looks like $\sum_{p_i}C_1(p_i)O_1(p_i)+C_2(p_i) O_2(p_i)$, we can use identity in \eq{identity} and reshuffle the momenta $p_i$  in the first term of \eq{identity} to make it look like the second term ($O_2$) with reshuffled momenta in the argument of the Wilson coefficient $C_1(p_1,p_2,p_3,p_4)\rightarrow C_1(p_3,p_2,p_1,p_4)$. As a result we reduced our basis of operators to only one: $O_2$ and Lagrangian takes form:
\begin{eqnarray}
\mathcal{L}_{\text{eff}}&=&-\lambda(\mu_h)\sum_{p_1,p_2,p_3,p_4}\, f(\mu_h, \mu_l, s, t, u)\,\phi^{\dagger}_4 \phi_3\phi^{\dagger}_2 \phi_1\nonumber\\
&=&-\lambda(\mu_h)\sum_{p_j}f(\mu_h,\mu_l,p_j)\sum_{i=1}^{4} R_{i2} \hat{O}_i(p_j).\label{LEFF}
\end{eqnarray}
There is one subtlety at this point, which is whether we set the hard scale $\mu_h=\sqrt{s}=\sqrt{(p_1+p_2)^2}$ before the summation in the Lagrangian, or we set this scale after taking the matrix element with external states.  We chose to do the later while the former would give a different numerical result. However the difference should be within hard scale variation, thus of the higher order, i.e. NNLL.

At NLL order the function $f$ is equal to:
\begin{eqnarray}
&&f_{\text{NLL}}(\mu_h,M_Z, s,t,u)=\e^{D(\mu_l=M_Z, s)}\nonumber\\
&&\times\Bigg(\Gamma_{22}(s,t,u)-\frac{1}{4}\Gamma_{12}(s,t,u)+\frac{1}{2}\Gamma_{12}(t,s,u)\Bigg).\nonumber\\\label{fNLL}
\end{eqnarray}
At LL it is independent of loop momenta completely:
\begin{eqnarray}
&&f_{\text{LL}}(\mu_h,\mu_l)=\Gamma_{\text{LL}}(\mu_h, \mu_l),\label{fLL}
\end{eqnarray}
where $\Gamma_{\text{LL}}$ is given in \eq{eq:GammaLL}.
\begin{figure*}[!t]
\includegraphics[width= 200pt]{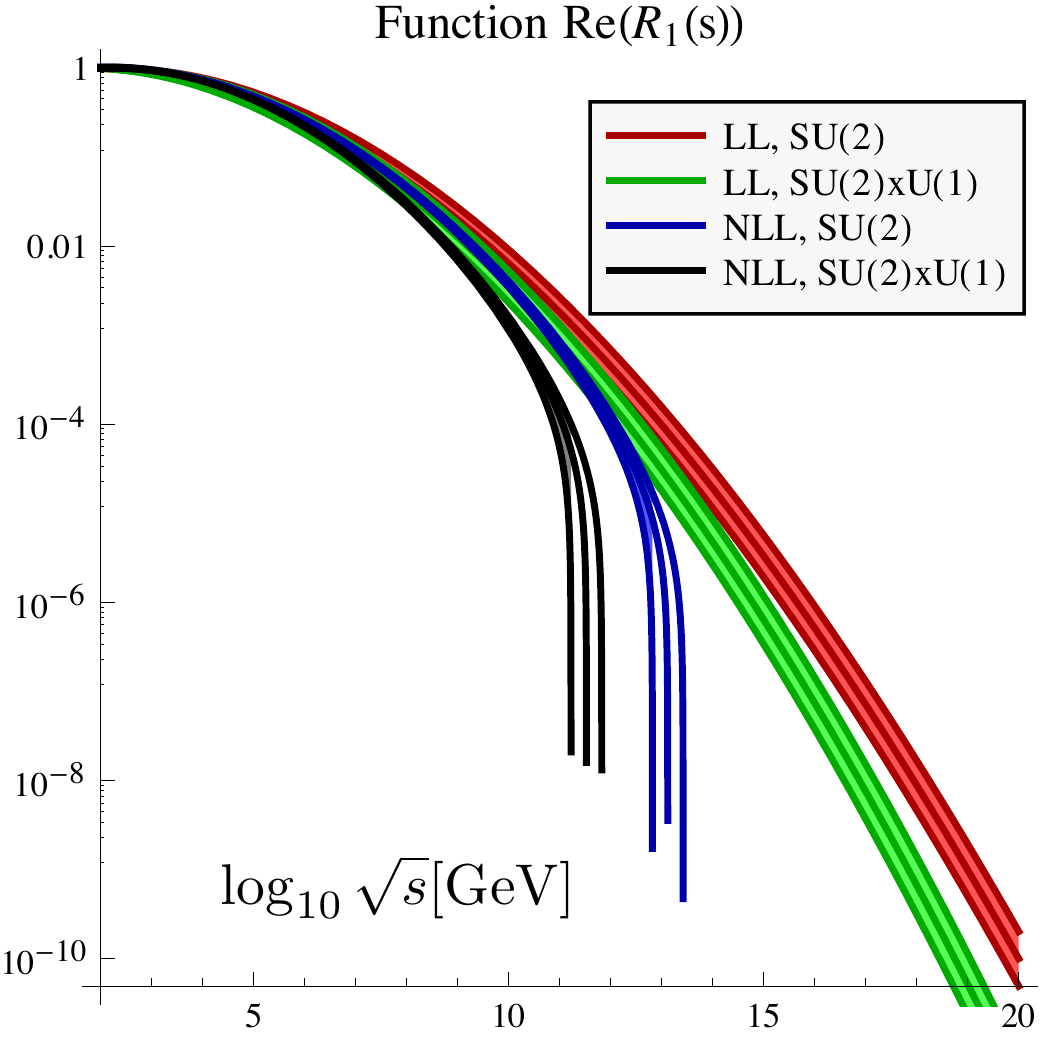} \hspace*{0.5in}  \includegraphics[width= 200pt]{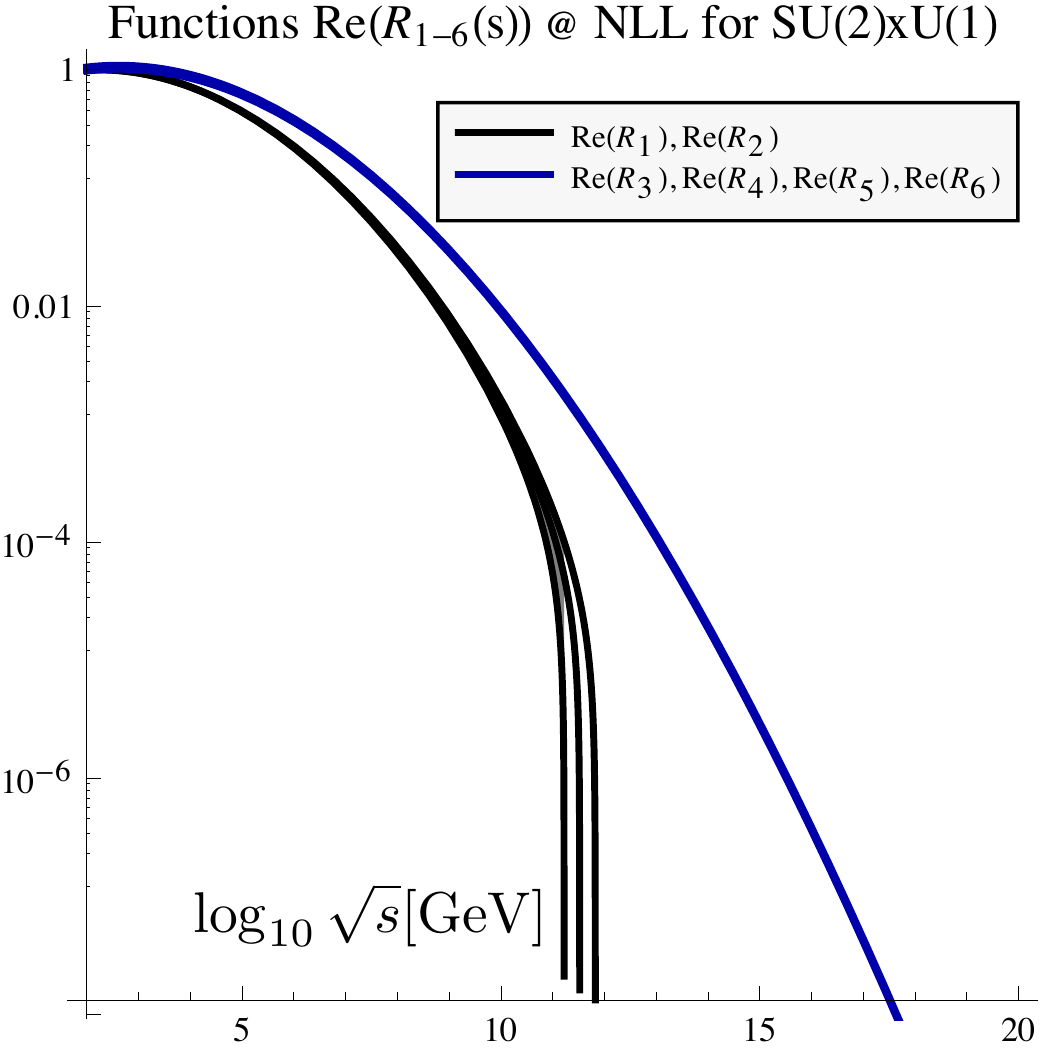}
\caption{Real part of $R_1(s)$ as function of center of mass energy (left) with illustrative breakdown of contributions from LL, NLL and SU(2) or combined SU(2)xU(1) (see the legend). Real part of all functions $R_i(s)$ at NLL for SM (right).}
\label{fig:numerics1}
\end{figure*}
\subsection{Evaluating $S-$matrix elements in the effective theory}
With effective theory Lagrangian found in the previous subsection it is a straightforward exercise to evaluate matrix elements of longitudinal gauge and Higgs boson scattering amplitudes. For example the amplitude for $W_LW_L\rightarrow W_L W_L$ scattering is given by:
\begin{eqnarray}
M_{W_LW_L\rightarrow W_L W_L}=\bra{w^+(\vc{k_3}),w^-(\vc{k}_4)}S\ket{w^+(\vc{k}_1) w^-(\vc{k}_2)},\nonumber\\
\end{eqnarray}
where this matrix element has to be evaluated with effective Lagrangian given in \eq{LEFF} and basis operators $\hat{O}_i$ in \eq{basis}. Defining the basis of normalized states\cite{Lee:1977eg} $e_i$:
\begin{eqnarray}
e_1&=&W_L^+(k_1)W_L^-(k_2), \qquad e_2=(1/\sqrt{2})Z_L(k_1) Z_L(k_2),\nonumber\\
e_3&=&(1/\sqrt{2})H(k_1)H(k_2), \qquad e_4=H(k_1) Z_L(k_2),
\end{eqnarray}
and similarly in the final state the basis states $e'_j$ can be found by substituting $k_1, k_2\rightarrow k_3, k_4$. The $S$ matrix elements of effective theory in this basis equal to:
\begin{eqnarray}
&&\bra{e'_j}S_{\text{EFT}}\ket{e_i}=-4\,\lambda(\mu_h)\nonumber\\
&&\times\left[ \begin{array}{cccc}
\frac{f_1+f_4}{2} &\frac{f_1+f_2}{4\sqrt{2}} & \frac{f_1+f_2}{4\sqrt{2}} & \frac{f_1-f_2}{4i}  \\
\frac{f_1+f_2}{4\sqrt{2}}&\frac{f_{\text{tot}}}{8} & \frac{f_{\text{tot}}-2(f_3+f_5)}{8} & 0  \\
\frac{f_1+f_2}{4\sqrt{2}} & \frac{f_{\text{tot}}-2(f_3+f_5)}{8}  & \frac{f_{\text{tot}}}{8} & 0  \\
\frac{f_1-f_2}{4i} & 0 & 0 & \frac{f_{\text{tot}}-2(f_1+f_4)}{4}  \end{array} \right],\nonumber\\\label{smatrix}
\end{eqnarray}
where $f_{\text{tot}}=\sum_{i} f_i$ and $f_i$ appear due to combinatorics because of different contractions. At tree level this leads to combinatorial factors, while in our case different contractions lead to different crossing of $s,t,u$ parameters:
\begin{eqnarray}
f_1&=&f(\mu_h,\mu_l, s,t,u), \qquad f_2=f(\mu_h,\mu_l, s,u,t),\nonumber\\
f_3&=&f(\mu_h,\mu_l, t,u,s), \qquad f_4=f(\mu_h,\mu_l, t,s,u),\nonumber\\
f_5&=&f(\mu_h,\mu_l, u,t,s), \qquad f_6=f(\mu_h,\mu_l, u,s,t),
\end{eqnarray}
and function $f$ is given to LL order in \eq{fLL} and to NLL order in \eq{fNLL}.
\begin{figure*}[!t]
\includegraphics[width= 200pt]{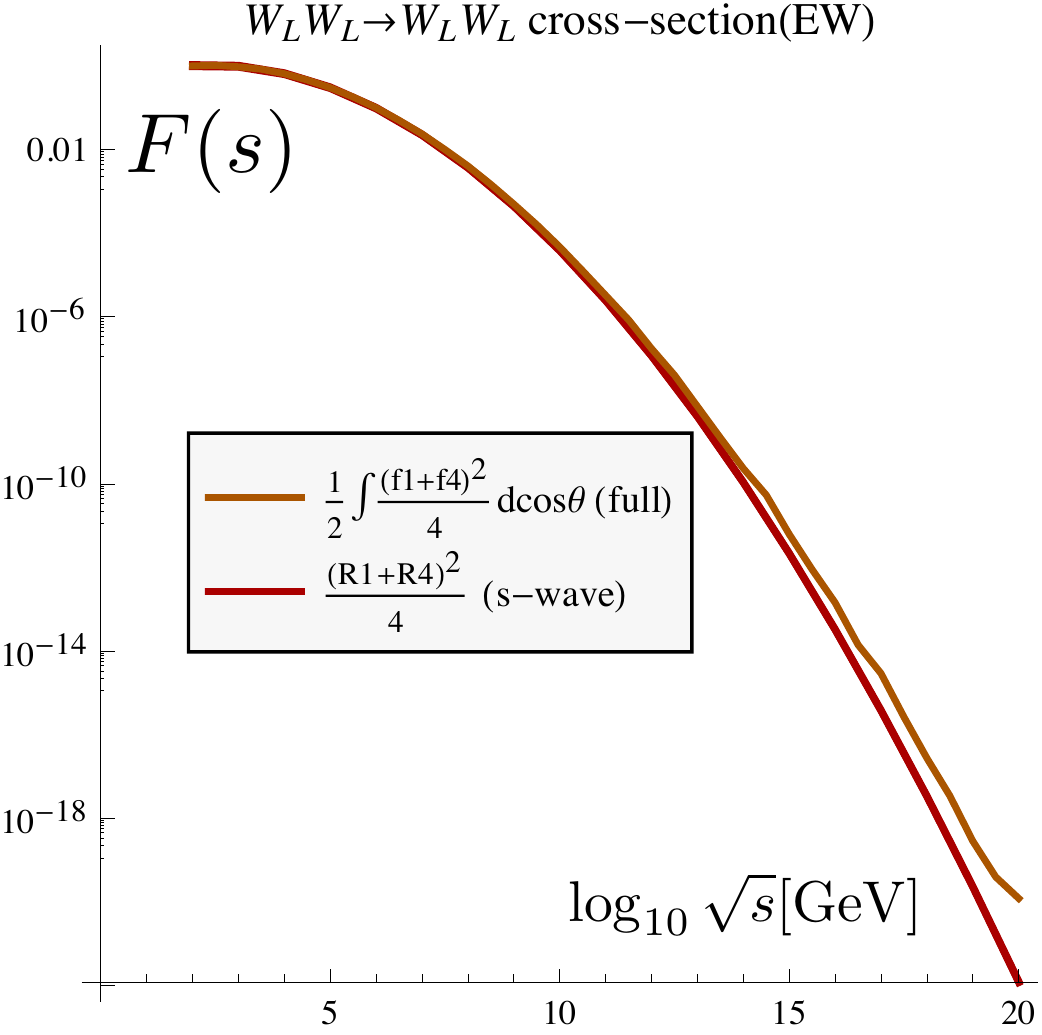} \hspace*{0.5in}  \includegraphics[width= 200pt]{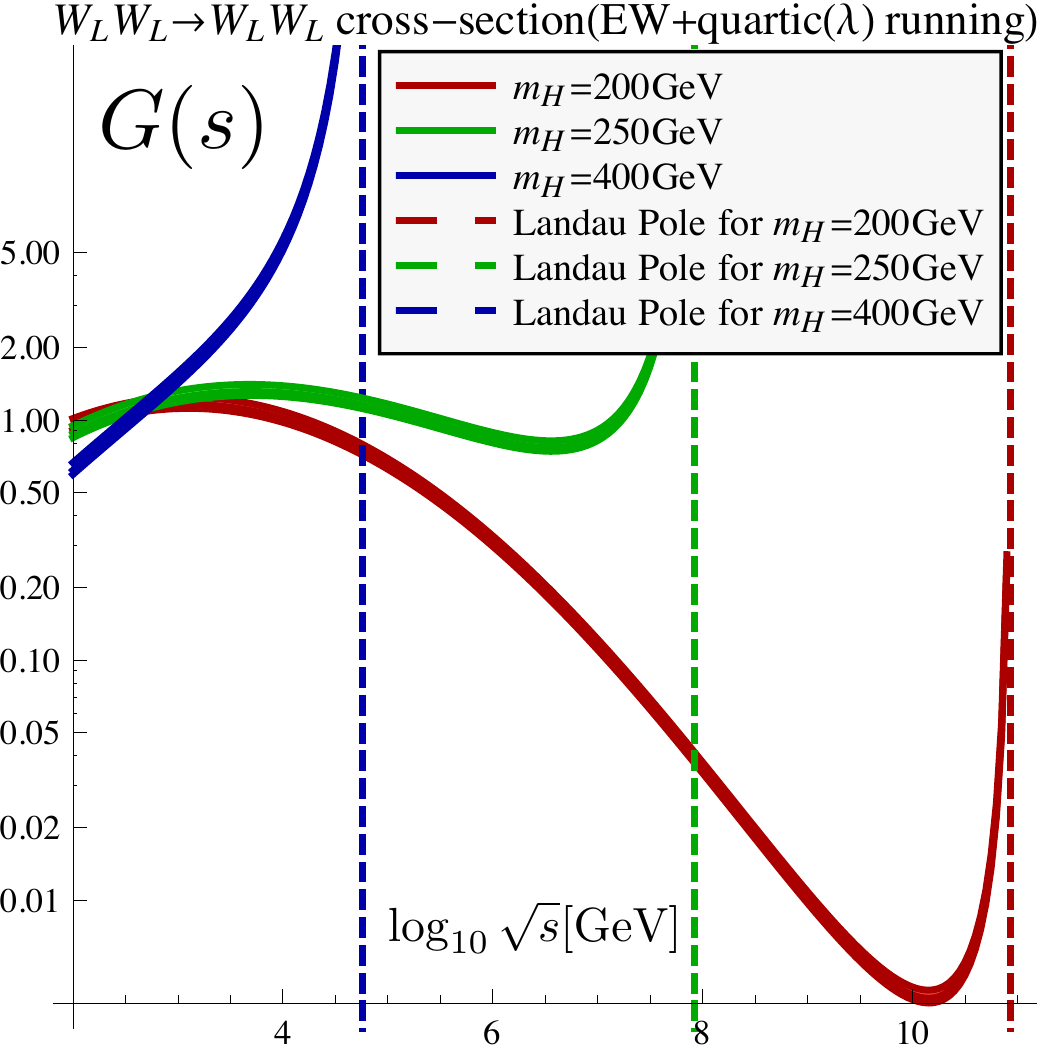}
\caption{Functions $F_1(s), F_2(s)$ defined in \eq{F1def} and \eq{F2def} (left plot) that control the electroweak part of the $W_{\text{L}}W_{\text{L}}$ cross section (see \eq{sigmaww}). Function $G(s)$ (right plot) that includes the Higgs quartic coupling and its running for three different Higgs masses.}
\label{fig:numerics2}
\end{figure*}
Note that at LL all functions $f_1-f_6$ are identical (due to our choice to set the scale $\mu_h$ after taking the matrix element of effective theory operators, as we discussed above), and we get that the $S$ matrix has a simple form:
\begin{eqnarray}
&&\bra{e'_j}S_{\text{EFT}}\ket{e_i}_{\text{LL}}=-4\,\lambda(\mu_h)\,\Gamma_{\text{LL}}\left[ \begin{array}{cccc}
1 &\frac{1}{\sqrt{8}} & \frac{1}{\sqrt{8}} & 0  \\
\frac{1}{\sqrt{8}}&\frac{3}{4} & \frac{1}{4} & 0  \\
\frac{1}{\sqrt{8}} & \frac{1}{4}  & \frac{3}{4} & 0  \\
0 & 0 & 0 & \frac{1}{2}  \end{array} \right].\nonumber\\\label{LLSmatrix}
\end{eqnarray}
Thus we find a simple result that at leading-logarithm order, including electroweak Sudakov logarithms, the entire tree level $S-$matrix\cite{Lee:1977eg} gets multiplied by a universal number $\Gamma_{\text{LL}}(\mu_h, \mu_l)$ given in \eq{eq:GammaLL}. To get numerical results one has to plug in $\mu_h=\sqrt{s}, \mu_l= M_{\text{EW}}\sim M_Z$ and scale variation can be used as usual as an estimate of theoretical uncertainties.

Also note that the LL result for the amplitude has no angular dependence, the scattering remains $s-$wave similarly to tree level result. The non-trivial angular dependence appears only at NLL order.

Finally, note that $\int (f_1(s,t,u)-f_2(s,t,u)) d\cos\theta=0$, which follows from $t(\cos\theta)=u(-\cos\theta)$. Thus in $s-$ wave analysis, the amplitude $W^+_\text{L}W^-_{\text{L}}\rightarrow H Z_L$ including electroweak Sudakov logarithms at NLL order vanishes. Same is true at tree level: amplitude $W^+_\text{L}W^-_{\text{L}}\rightarrow H Z_L$  {\it{only vanishes in}} $s-$wave and $Z_{\text{L}}Z_{\text{L}}\rightarrow H Z_L$ and $HH\rightarrow H Z_L$ {\it{vanish identically}}\cite{Lee:1977eg}.
\subsection{Numerical application to $W^+_LW^-_L\rightarrow W^+_L W^-_L$ cross section}
In this subsection we will use the general $S-$ matrix derived in the previous subsection at NLL order to numerically study the NLL order electroweak Sudakov logarithms in $W_L^+W_L^-\rightarrow W_L^+W_L^-$ cross section. This process does not provide the tightest bound for traditional unitarity bound analysis of Standard Model \cite{Lee:1977eg, Marciano:1989ns, Kolda:2000wi}. Nevertheless we provide this as an illustration to the effect of electroweak logarithms for longitudinal gauge boson scattering cross sections. As was found in the previous subsection at LL order all other processes are proportional to each other, i.e. to the one considered in this subsection, the only non-trivial difference appears at NLL order.

Since at LL $s-$wave analysis is exact, at NLL order we expect it to be a good approximation. Thus, we define functions $R_i$ which are given by:
\begin{eqnarray}
R_i(s)=\frac{1}{2}\int f_i(s, \cos\theta) d\cos\theta.
\end{eqnarray}
How this functions depend on energy controls the $s-$wave analysis of cross-section with longitudinal gauge and Higgs bosons. In Figure \ref{fig:numerics1} we present this dependence for all functions $\text{Re} R_{i}$. The left of this figure is function $\text{Re} R_1(s)$ with breakdown of contributions from separately LL SU(2) only Sudakov corrections(red), LL SU(2)xU(1) Sudakov corrections(green) and NLL SU(2) (blue) and finally full NLL SU(2)xU(1) Sudakov corrections (black). The right plot in Figure \ref{fig:numerics1} represents the full NLL SU(2)xU(1) Sudakov corrections for each function $\text{Re} R_{1-6}$. It turns out that to this order there are only two non-trivial functions: $R_1=R_2$ and $R_3=R_4=R_5=R_6$ as shown in that figure. The reason $R_1=R_2$ has been explained at the end of previous subsection. By the same argument it is clear that $R_3=R_5$ and $R_4=R_6$. Numerically it so happens that they all are the same at this order, which could be explained if one completely ignores the angular dependence in $f_i$. The horizontal axis for $\sqrt{s}$ goes all the way to the Planck scale. As expected at low energies the Sudakov corrections are small, however very quickly they reduce amplitude at tree level by a factor of $10^{-4}$ at the CM energy $10^{12}\text{GeV}$.

Having studied the generic behavior of scattering amplitudes of longitudinal gauge bosons (and Higgs), we specialize to the case of $W^+_\text{L}W^-_\text{L}\rightarrow W^+_\text{L} W^-_\text{L}$:
\begin{eqnarray}
\sigma\left(W^+_\text{L}W^-_\text{L}\rightarrow W^+_\text{L} W^-_\text{L}\right)=\sigma_{\text{tree}} \left(\frac{\lambda (\sqrt{s},m_H)}{\lambda (m_H,m_H)}\right)^2 F(s),\nonumber\\\label{sigmaww}
\end{eqnarray}
where function $F(s)$ we calculate by two methods. First is exact, keeping the full angular dependence in the amplitude, coming from NLL order (see \eq{smatrix}):
\begin{eqnarray}
F_1(s)=\frac{1}{2}\int d\cos\theta \frac{|f_1+f_4|^2}{4}.\label{F1def}
\end{eqnarray}
In the second method we use $s-$wave approximation:
\begin{eqnarray}
F_2(s)=\frac{|R_1+R_4|^2}{4}.\label{F2def}
\end{eqnarray}
Both functions are plotted in the left part of Figure \ref{fig:numerics2}. It is comforting to see, that there is a good agreement between the exact and $s-$wave approximated cross sections, since at leading logarithmic order the entire amplitude is a pure $s-$wave. While it is interesting that the electroweak corrections reduce the $WW$ scattering cross section so dramatically at high energies, $10^{-10}$ at the CM energy of $10^{13}\text{GeV}$, one should keep in mind that depending on the Higgs mass $m_H$, the quartic coupling $\lambda(\mu)$ has a Landau pole and extending predictions beyond that scale is meaningless. As an illustration, on the right part of Figure \ref{fig:numerics2} we include the Higgs coupling and plot the function $G(s)=\lambda^2(\sqrt{s},m_H)/\lambda^2(m_H,m_H)F_1(s)$ for different Higgs masses. As expected the cross section blows up at the Landau pole, which is unaffected by Sudakov corrections. However for lighter Higgs masses ($<400\text{GeV}$), almost everywhere except the vicinity of the Landau pole, the cross section is significantly suppressed due to Sudakov double logarithms.
\section{Applications for unitarity analysis}\label{seq:Unitarity}
The standard analysis of unitarity bound of SM \cite{Lee:1977eg, Marciano:1989ns, Kolda:2000wi} includes longitudinal gauge boson scattering amplitudes at tree level \cite{Lee:1977eg} and in Refs. \cite{Marciano:1989ns, Kolda:2000wi} the running of the Higgs quartic coupling $\lambda(\mu)$ is included as well as improved unitarity bound $\text{Re}(a_0)<1/2$, where $a_0$ is the $s-$partial wave of the scattering amplitude $M(s,\cos\theta)$. General expansion for the amplitude in terms of spherical harmonics is given by:
\begin{eqnarray}
M(s,\cos\theta)=16\pi\sum_0^{\infty}(2J+1)a_J(s)P_J(\cos\theta).\label{mexpand}
\end{eqnarray}
Inverting \eq{mexpand} we find:
\begin{eqnarray}
a_J=\frac{1}{32\pi}\int_{-1}^1 M(s,\cos\theta) P_J (\cos\theta) d\cos\theta.\label{aJformula}
\end{eqnarray}
Taking into account the resummation of electroweak logarithms as derived in section \ref{seq:WW} leads at NLL order to some non-trivial $p, d,$etc partial waves, due to angular dependence of $M(s, \cos\theta)$, while at LL such dependence is absent. At tree level $f_i=1$ and plugging elements of \eq{smatrix} into \eq{aJformula} for $J=0$ we reproduce correctly the tree level result\cite{Lee:1977eg}:
\begin{eqnarray}
(a_0)_{\text{tree}}=-\frac{\lambda(\sqrt{s})}{4\pi} \left[ \begin{array}{cccc}
1 & \frac{1}{\sqrt{8}}&\frac{1}{\sqrt{8}}& 0  \\
\frac{1}{\sqrt{8}} & \frac{3}{4}& \frac{1}4& 0  \\
\frac{1}{\sqrt{8}} & \frac{1}{4}& \frac{3}{4}& 0  \\
0 & 0& 0& \frac{1}{2}
 \end{array} \right].
\end{eqnarray}
\begin{figure}[!t]
\includegraphics[width= 200pt]{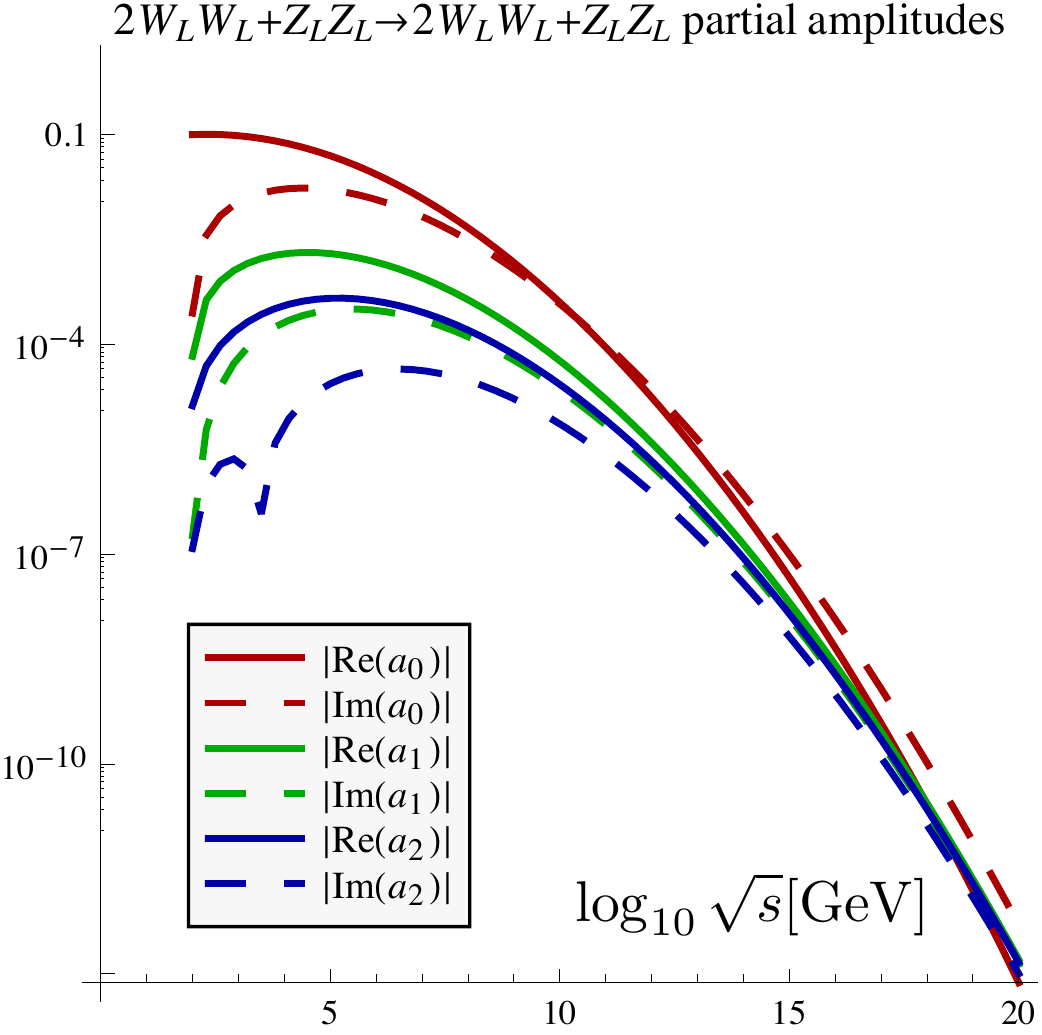} 
\caption{Absolute values of real and imaginary parts of first three partial waves for  $(2W^+_\text{L}W^-_{\text{L}}+Z_{\text{L}} Z_{\text{L}})/\sqrt{6}$ scattering. Note that at high energies higher partial waves catch up with $\text{Re} a_0$.}
\label{fig:numerics3}
\end{figure}
The tightest bound for unitarity analysis is given by eigenvector of this matrix corresponding to largest eigenvalue. The eigenvalues in the units of $-\lambda/4\pi$ are $3/2,1/2,1/2,1/2$. The largest eigenvalue corresponds to eigenchannel $(2W_{\text{L}}^+W_{\text{L}}^-+Z_{\text{L}}Z_{\text{L}}+HH)/\sqrt{8}$. However in \cite{Marciano:1989ns} this state is not used for unitarity bound, since for heavy enough Higgs mass the scale $\Lambda$ where unitarity of SM breaks  becomes close to Higgs mass, thus the state under consideration is kinematically not allowed. Instead in \cite{Marciano:1989ns} the isospin zero state $(2W^+_\text{L}W^-_{\text{L}}+Z_{\text{L}} Z_{\text{L}})/\sqrt{6}$ is used and the unitarity bound is a factor of $5/6$ weaker for this state however the problem mentioned above is absent.

Now our goal is to re-derive the scale where SM unitarity is broken by extending it from tree level, to including the Sudakov double logarithms that we have calculated in Section \ref{seq:WW} for arbitrary longitudinal gauge or Higgs boson scattering to NLL order . Since we found that at LL, there is a simple multiplicative factor for $S-$ matrix (see \eq{LLSmatrix}) we can simply make the same choice of the state $(2W^+_\text{L}W^-_{\text{L}}+Z_{\text{L}} Z_{\text{L}})/\sqrt{6}$ as in \cite{Marciano:1989ns} as our candidate to give the best unitarity bound. This is not true at NLL, but corrections should be small. Also it is an apple-to-apple comparison to \cite{Marciano:1989ns}, where the tree level amplitude and one-loop Higgs quartic running have been included into analysis. We limit ourself to unitarity bound $\text{Re} a_0<1/2$ as in \cite{Marciano:1989ns}, although at very high energies as we will see below the imaginary parts and higher harmonics start to play role due to NLL effects.

In order to understand the role of higher harmonics $a_1, a_2, etc.$ which appear due to NLL order Sudakov double logarithms, as well as the imaginary parts of scattering partial waves, we plotted in Figure \ref{fig:numerics3} the real and imaginary parts of first three harmonics: $a_0, a_1, a_2$ as function of $\sqrt{s}$ specifically for the $(2W^+_\text{L}W^-_{\text{L}}+Z_{\text{L}} Z_{\text{L}})/\sqrt{6}$ channel, which is a simple exercise given \eq{smatrix} and \eq{aJformula}. For energies around $\sqrt{s}=10^{16}\text{GeV}$ higher partial waves become comparable to $a_0$, although still smaller. Also for $\sqrt{s}>10^{10}\text{GeV}$ the imaginary part of $a_0$ becomes bigger than the real part. 

\begin{figure}[!t]
\includegraphics[width= 200pt]{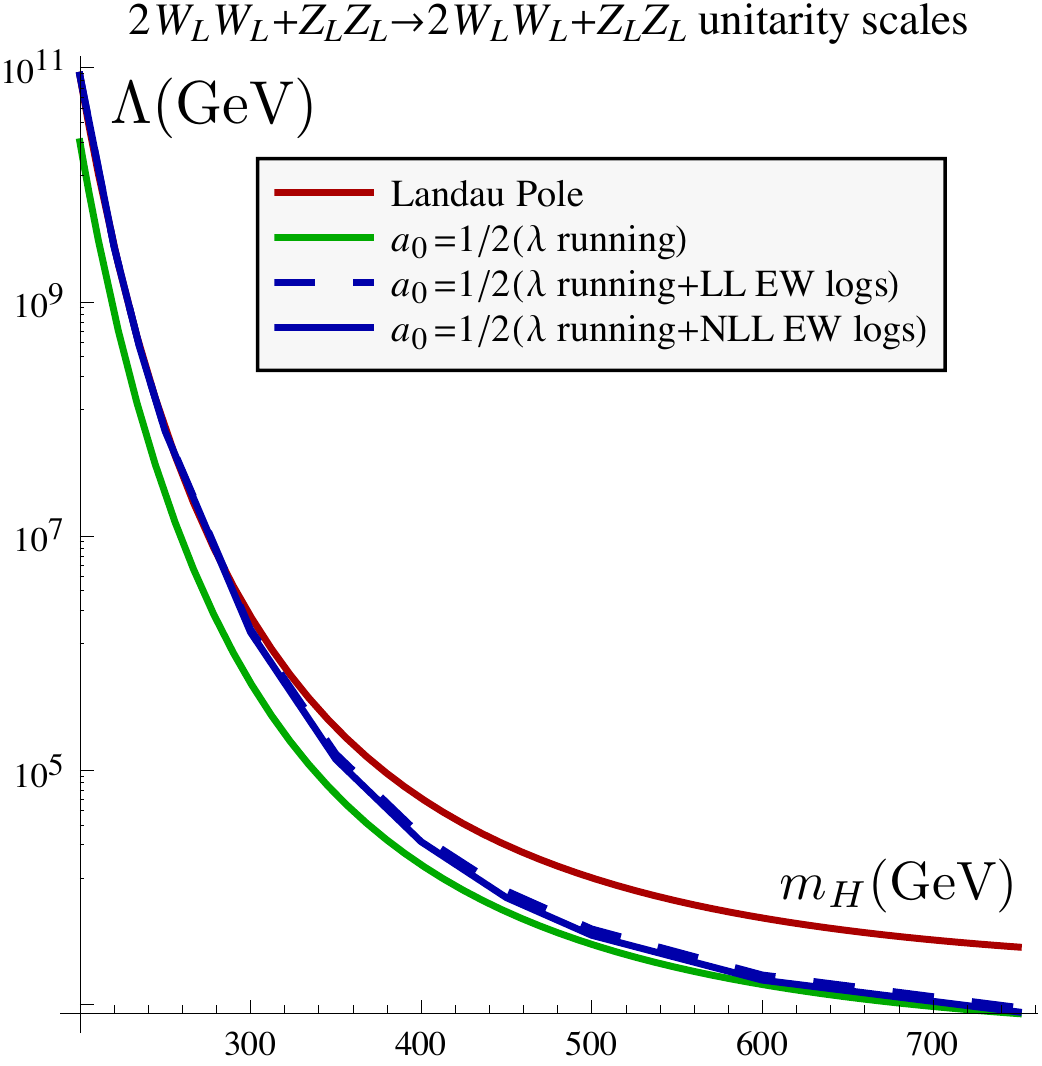}
\caption{Unitarity bound of SM. Red line corresponds to the position of the Landau pole. Green line corresponds to the scale where $a_0=1/2$ for scattering channel $(2W^+_\text{L}W^-_{\text{L}}+Z_{\text{L}} Z_{\text{L}})/\sqrt{6}$  at tree level analysis of  \cite{Marciano:1989ns} that includes the running of Higgs self-coupling. The dashed blue and blue lines corresponds to bound $\text{Re}a_0=1/2$ including electroweak Sudakov logarithms at LL and NLL correspondingly.}
\label{fig:numerics4}
\end{figure}

We turn to the unitarity bound of Standard Model including the electroweak Sudakov resummation. In Figure \ref{fig:numerics4} we present our results for unitarity bounds on $(2W^+_\text{L}W^-_{\text{L}}+Z_{\text{L}} Z_{\text{L}})/\sqrt{6}$ scattering. The red curve is simply the position of Landau pole as a function of Higgs mass. We use one-loop beta function for quartic coupling including only self coupling contribution to the beta function, which is a good approximation for heavier Higgs masses. The green curve in Figure \ref{fig:numerics4} represents the unitarity bound found in \cite{Marciano:1989ns}, i.e. requiring $\text{Re}a_0=1/2$ and including Higgs quartic coupling one loop running. Including electroweak logarithms into the formula for the amplitude we get the dashed blue curve (LL) and solid blue curve (NLL). 

\begin{figure*}[!t]
\includegraphics[width= 200pt]{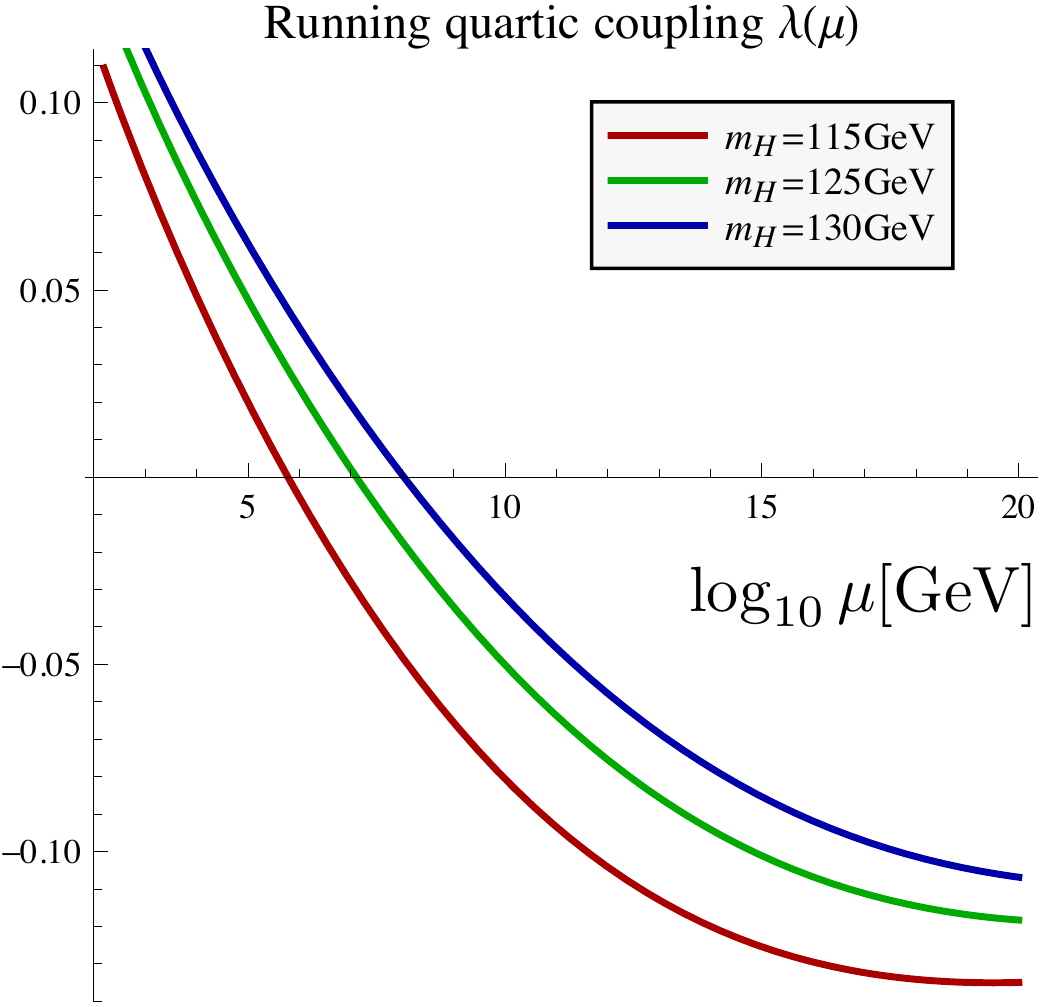} \hspace*{0.5in}  \includegraphics[width= 200pt]{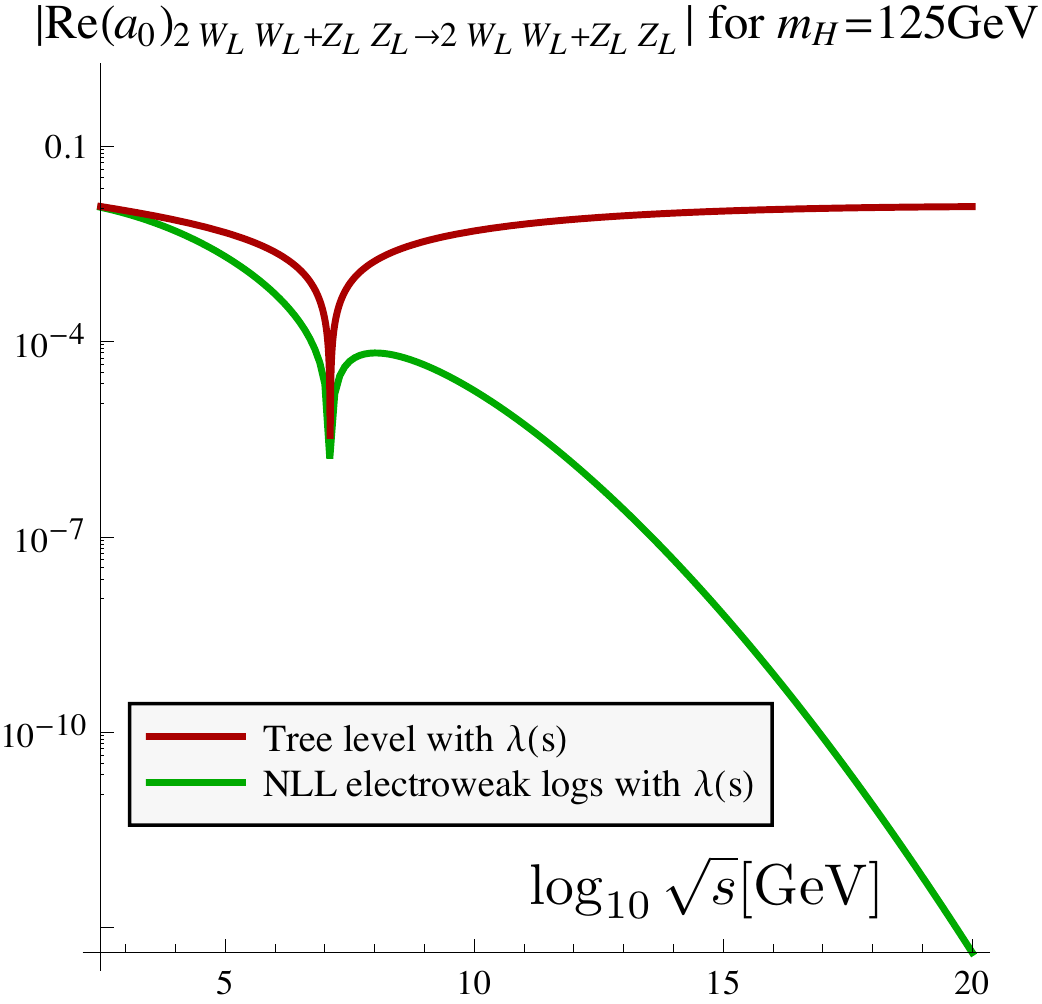}
\caption{Running quartic coupling for Higgs masses 115, 125 and 130\text{GeV} correspondingly (left). Real part of partial wave for unitarity constraining scattering of the state $(2W_{\text{L}}^+W_{\text{L}}^-+Z_{\text{L}}Z_{\text{L}})/\sqrt{6}$ for $m_H=125\text{GeV}$ (right).}
\label{fig:numerics5}
\end{figure*}

Finally, let us address the role of electroweak logarithms and unitarity scales for the case of light Higgs mass which is favorable from current LHC bounds. Simple formula in \eq{quarticrunning} considered above is not applicable in this case. In Ref.\cite{Arason:1991ic} beta functions of all couplings of SM are computed to two-loop order\footnote{I thank Aneesh Manohar for pointing out that reference.}. For simplicity we only use their results constraining ourselves to one-loop, including running of couplings of SM: $g', g, g_3, \lambda, y_t$ and ignoring remaining Yukawa couplings. Note the different normalization(factor of 2) between \cite{Arason:1991ic} and us for $\lambda$, and $g_{1}$ in \cite{Arason:1991ic} is same as $g'\sqrt{5/3}$, while $g_2, g_3$ are same as our $g, g_3$ and $\alpha_1=g'^2/4\pi, \alpha_2=g^2/4\pi$ and $\alpha_s=g_3^2/4\pi$. As a result we get the left plot in Figure \ref{fig:numerics5} for Higgs masses $m_H=115,125, 130\text{GeV}$. As one can see the qualitative behavior is very much different than for heavy Higgs masses. Instead of Landau pole, we get a fixed point at high energies as well as the change in the sign of the quartic coupling at an energy scale that depends on the mass of the Higgs boson. Similar behavior of quartic coupling was found for Higgs mass around $m_H=125\text{GeV}$ in Ref.\cite{EliasMiro:2011aa}, which used state-of-the art analysis, including full two-loop running of all couplings. 

As a result for light Higgs masses shown in the left part of Figure \ref{fig:numerics5} we see that the coupling remains perturbative to Planck scale. Furthermore, as one can see from the red plot in the right part of Figure \ref{fig:numerics5}, the tree level $s-$wave partial amplitude for unitarity constraining scattering of the state $(2W^+_\text{L}W^-_{\text{L}}+Z_{\text{L}} Z_{\text{L}})/\sqrt{6}$ is within the unitary limit $Re a_0<1/2$ for all center of mass energies all the way to the Planck scale. The electroweak logarithms modify this amplitude by reducing it significantly  at high energies. For example at $\sqrt{s}=10^{14}\text{GeV}$ the suppression factor is of the order $10^{-5}$. Thus for phenomenologically interesting low Higgs masses, the unitarity is preserved at all scales. The role of electroweak logarithms is suppression of the amplitude (and cross section) at high energies compared to tree level result.

\section{Conclusions}\label{seq:concl}
We have computed to next-to-leading logarithmic order the effect of electroweak Sudakov logarithms for any longitudinal gauge or Higgs boson scattering amplitude. At leading-logarithmic order the amplitude gets multiplied by a universal function $f_{\text{LL}}$, independent of the concrete process and with strong dependence on energy. For example, the high energy behavior of the function $f$ for $W^+_{\text{L}}W^-_{\text{L}}\rightarrow W^+_{\text{L}}W^-_{\text{L}}$reduces the tree level cross section suppressed by a factor ranging from $\approx 0.3$ at $\sqrt{s}=100\text{TeV}$ to $\approx 10^{-12}$ at $\sqrt{s}=10^{13}\text{TeV}$.  This correction does not depend (at our order) on Higgs mass. Of course the entire approach becomes intractable for energies above the Landau pole of the quartic coupling $\lambda$.

This kind of  behavior at high energies affects the unitarity bound of SM. Despite the Sudakov suppression, when energy gets close to quartic Landau pole, the singularity still wins in the  very vicinity of the pole and the unitarity is still broken. 

For heavier Higgs masses, the Landau pole appears very close to the Higgs mass, so the electroweak logarithms do not get a chance to contribute significantly, since at those scales the logarithms are small. In this limit the unitarity bound is unaffected. However for lighter Higgs masses, (but still $>\text{200}\text{GeV}$)when the Landau pole gets further and further from the electroweak scale, the electroweak corrections play a dramatic role for the high energy behavior of the longitudinal gauge boson scattering, by pushing the scale where the unitarity of SM is broken very close to the Landau pole energy scale, thus restoring unitarity except in the immediate vicinity of the pole. This qualitative picture can be easily seen in figure \ref{fig:numerics4}.

Finally for very light Higgs mass around $125\text{GeV}$, which is currently phenomenologically favorable, the quartic coupling remains perturbative at all scales and preserves unitarity at all scales, both at tree level and when electroweak logarithms are included, which suppress amplitude and cross section of longitudinal gauge boson scattering significantly at high energies. This can be seen from Figure \ref{fig:numerics5}. 


\begin{acknowledgments}
I would like to thank Vincenzo Cirigliano, Alexander Friedland, Terry Goldman and Michael Graesser for useful discussions. I also thank Aneesh Manohar for helpful correspondence about the draft of this paper. Special thanks to Olga Serafimova for inspiring me to complete this paper. This research is  supported by the US Department of Energy, Office  of Science, under
Contract No. DE-AC52-06NA25396 and in part by the LDRD program at LANL and the JET topical 
collaboration. 
\end{acknowledgments}

\end{document}